\documentclass[hyphens]{cai}
\usepackage{graphicx}

 \volume{32}
 \yyear{2013}
 \page{1001}

\usepackage[utf8]{inputenc}

\usepackage{hyperref}
\usepackage[super]{nth}
\usepackage{amsmath}%
\usepackage{amssymb}%
\usepackage{wasysym}%
\usepackage{adjustbox}
\usepackage{listings}
\usepackage[inline]{enumitem}
\usepackage[frozencache=true]{minted}
\usepackage{mdframed}
\usepackage{multirow}
\usepackage{graphicx} 

\newcommand{\thref}[1]{\textbf{\hyperref[#1]{RQ~\ref{#1}}}}
\newtheorem{theorem}{RQ}

\newcommand{\hypref}[1]{\textbf{\hyperref[#1]{H~\ref{#1}}}}
\newtheorem{hyp}{H}

\begin{document}
\label{firstpage}

\title[Test Case Identification in Java OS Projects on GitHub]
      {Automating Test Case Identification in Java Open Source Projects on GitHub}

\author[M. Madeja, J. Porubän, M. Bačíková]
       {Matej \surname{Madeja}, Jaroslav \surname{Porubän}, Michaela \surname{Bačíková}}
\affiliation{Department of Computers and Informatics\\
Faculty of Electrical Engineering and Informatics\\
Technical University of Košice, Letná 9, 042 00 Košice, Slovakia}
\email{\{matej.madeja\}\{jaroslav.poruban\}\{michaela.bacikova\}@tuke.sk}

\author[M. Sulír, J. Juhár, S. Chodarev, F. Gurbáľ]
      {Matúš \surname{Sulír}, Ján \surname{Juhár}, Sergej \surname{Chodarev}, Filip \surname{Gurbáľ}}
\affiliation{Department of Computers and Informatics\\
Faculty of Electrical Engineering and Informatics\\
Technical University of Košice, Letná 9, 042 00 Košice, Slovakia}
\email{\{matus.sulir\}\{jan.juhar\}\{sergej.chodarev\}\{filip.gurbal\}@tuke.sk}

%
%

\noreceived{} \nocommunicated{}

\maketitle

\begin{abstract}
Software testing is one of the very important Quality Assurance (QA) components. A lot of researchers deal with the testing process in terms of tester motivation and how tests should or should not be written. However, it is not known from the recommendations how the tests are written in real projects. In this paper, the following was investigated:  (i) the denotation of the word ``test'' in different natural languages; (ii) whether the number of occurrences of the word ``test'' correlates with the number of test cases; and (iii) what testing frameworks are mostly used. The analysis was performed on 38 GitHub open source repositories thoroughly selected from the set of 4.3M GitHub projects. We analyzed 20,340 test cases in 803 classes manually and 170k classes using an automated approach. The results show that: (i) there exists a weak correlation ($r = 0.655$) between the number of occurrences of the word ``test'' and the number of test cases in a class; (ii) the proposed algorithm using static file analysis correctly detected 97\% of test cases; (iii) 15\% of the analyzed classes used \texttt{main()} function whose represent regular Java programs that test the production code without using any third-party framework. The identification of such tests is very complex due to implementation diversity. The results may be leveraged to more quickly identify and locate test cases in a repository, to understand practices in customized testing solutions, and to mine tests to improve program comprehension in the future.
\end{abstract}


\begin{mathclass}
68-04
\end{mathclass}

\newcommand{\hACorrelation}{There is a strong correlation ($r \notin (-0.8, 0.8)$) between the number of occurrences of the word \textit{``test''} in the file content and the number of test cases.}
\newcommand{\rqAFrameworks}{How many testing classes are implemented as customized testing solutions without using any \nth{3} party framework?}

\newcommand{\hACorrelationNull}{There is \textbf{not} a strong correlation ($r \in (-0.8, 0.8)$) between the number of occurrences of the word \textit{``test''} in the file content and the number of test cases in projects with high number of ``test'' occurrence.}


\section{Introduction}
\label{sec:intro}
The development of automated tests in a software project is a time-consuming and costly process, as it represents more than half of the entire development process~\cite{7503707}. The main aim of testing is to maintain the quality of the product and in addition to that tests describe the expected behavior of the production code being tested. Years ago, Demeyer et al.~\cite{demeyer2002object} suggested that if the tests are maintained together with the production code, their implementation is the most accurate mirror of the product specification and can be considered as up-to-date documentation. Tests can contain many useful production code metadata that can support program comprehension.

Understanding the code is one of the very first tasks a developer must struggle with before the implementation of a particular feature. When the product specification changes (e.g. the requirements for new features are added), the developer must first understand them, then create his/her mental model~\cite{CORRITORE199961} and finally, the created mental model is expressed in a specific artifact --- code implementation. The problem is that two developers are likely to create two different mental models for the same issue because according to Mayer~\cite{10.1145/356835.356841} mental model may vary with respect to its completeness and veridicality. A comprehension gap could arise when one developer needs to adapt another programmer's mental model from the code.

An assumption can be made that by using the knowledge about the structure and semantics of tests and their connection to the production code, it is possible to increase the effectiveness of program comprehension and reduce the comprehension gap. This would be possible, for example, by enriching the source code with metadata from the tests directly into the production code, e.g. data used for testing, test scenarios, objects relations, comments, etc. To achieve this goal, it is necessary to know in detail how the tests are actually written and what data they use.

There exist many guidelines on how tests should be created. First, naming conventions may aid the readability and comprehension of the code. According to the empirical study by Butler et al.~\cite{7332450}, developers largely follow naming conventions. 
Our previous research~\cite{madeja_et_al:OASIcs:2019:10870} shows that there is a relation between the naming of identifiers in the test code and the production code being tested. This indicates that the relationship between the test and production code is not only at the level of method calls, object instances, or identifier references, but also at the vocabulary level, depending on the domain knowledge and mental model of a tester/developer.

Furthermore, many authors~\cite{nayyar2019instant,lewis2017software,garcia2017mastering} define best practices to simplify the test with the benefit of a faster understanding of the testing code and the identification of test failure. Some guidelines lead to avoiding test smells~\cite{van2001refactoring} because as reported by recent studies~\cite{peruma2019distribution,8529832}, their presence might not only negatively affect the comprehension of test suites but can also lead to test cases being less effective in finding bugs in the production code. All mentioned approaches are only recommendations but do not really express how the tests are written in real projects. That means we know how tests should be written, but we do not know how they are written in practice. Many researchers have tried to clarify the motivation of writing tests~\cite{8094467,10.1145/2786805.2786843,7102609}, the impact of test-driven development (TDD) on code quality~\cite{7592412,BISSI201645} or the popularity of testing frameworks~\cite{7884645}.

To reveal testing practices in real and independent projects it is necessary to find a way to identify test cases in a project, without the time-consuming code analysis. Much more important than the number of test cases is the information where they are located. When a testing framework is used, the test identification is mostly straightforward, e.g. by the presence of the framework imports. On the other hand, to obtain a general overview of testing practices regardless of the used framework, it is advisable to consider tests that do not use any third-party framework and can be regarded as customized testing solutions. In most of the related works, tests are identified by searching specific file and folder names, or some specific keywords. Considering that these keywords usually included the word ``test'' and based on the authors' experience of Java test cases development, it can be assumed that there is a relation between the word ``test'' and the number of test cases in a file. That means searching for the ``test'' string could be beneficial for faster test case identification. Based on the previous reasoning, this paper defines the following hypothesis and research question:

\begin{hyp}\label{h:hACorrelation}\hACorrelation\end{hyp}
\begin{theorem}\label{rq:rqAFrameworks}\rqAFrameworks\end{theorem}

This paper is focused exclusively on unit testing and analyzes 38 projects that have been carefully selected (see section~\ref{sub:relevant-projects}) from all GitHub projects with Java as a primary language (most of the code written in Java). Section \ref{sec:bg} presents the current state and found gaps in the research. In section \ref{sec:method}, the research method is described, containing an examination of whether it is appropriate to search for tests using the word ``test'' due to different natural languages of developers, an overview of known testing frameworks, and a proposed algorithm for static code analysis to automate the identification of test cases. Section \ref{sec:results} summarizes the results, threats to validity are mentioned in section \ref{sec:threats}, and conclusions can be found in section \ref{sec:concl}.


\section{State of the Art}
\label{sec:bg}
Many researchers examine software testing but we still know little about the structure and semantics of test code. This chapter summarizes the related work of software testing from various perspectives.

Learning about real testing practices is a constant research challenge. The goal of such research is mostly to find imperfections and risks, learn, and make recommendations on how to prevent them and how to streamline their development. Leitner and Bezemer~\cite{10.1145/3030207.3030213} studied 111 Java-based projects from GitHub that contain performance tests. Authors identify tests by searching for one or more terms in the test file name or for the presence of popular framework import, solely in the \texttt{src/test} project directory. Selected projects were subjected to manual analysis, in which they monitored several metrics. The most important result for this paper was the fact that 103 projects also included unit tests, usually following standardized best practices. On the other hand, the performance testing approach of the same projects often appears less extensive and less standardized. Another finding was that 58 projects (52\%) mix performance tests freely with their functional test suite, i.e., performance tests are in the same package, or even the same test file, as functional tests. Six projects implemented tests as usage examples. Using a similar approach~\cite{10.1145/3030207.3030213}, in our case by searching for the word ``test'' and searching for imports of testing frameworks in all project's Java files, we would like to analyze unit tests, but with a careful selection from all GitHub projects at a specific time, resulting in more relevant projects used for analysis.

Code coverage, also known as test coverage, is a very popular method for evaluating project quality. Ellims et al.~\cite{1383101} investigated the usage of unit testing in practice in three projects that authors evaluated as well-tested. Statement coverage was found to be indeed a poor measure of test adequacy. According to the findings of Hemmati~\cite{7272926}, basic criteria such as statement coverage are a very weak metric, detecting only 10\% of the faults. A test case may cover a piece of code but miss its faults. According to Hilton et al.~\cite{10.1145/3238147.3238183}, coverage can be beneficial in the code review process if a smaller part of the project is evaluated. By reducing coverage to a single ratio of the whole project, much valuable information could be lost. Kochhar et al.~\cite{8031982} performed an analysis of 100 large open-source Java projects showing that 31\% of the projects have coverage greater than 50\% and only 8\% are greater than 75\%.

Many experiments try to express the quality of tests by testing ``mutants''~\cite{10.1145/2635868.2635929}, i.e., by modifying a program in small ways to create artificial defects. According to Gopinath et al.~\cite{6982626} mutants do not necessarily represent real bugs, therefore, they are not able to relevantly evaluate the quality of the test suite nor to find relations between the coverage and mutants' reveal. However, there is a statistically significant correlation between code coverage and bug kill effectiveness of real software errors (non-mutants)~\cite{7081877}. The quality of the test suite is influenced by the way the mental model is expressed in the code, so examining real tests is more beneficial instead of using mutants.

The fact that unit tests are the most common test type in a project is confirmed by Cruz et al.~\cite{Cruz2019}: 39\% of 1000 analyzed Android projects used unit tests. Another finding was that frequently updated projects were more aware of the importance of using automated tests than those updated several years ago. The adoption of tests has increased over the last few years, so focusing on information mining from the tests makes sense.

Another type of research was done by  Munaiah et al.~\cite{Munaiah2017}, who focused on the assessment of GitHub projects. They proposed a tool that can be used to identify repositories containing real engineered software projects. The aim was to eliminate the repository noise such as example projects, homework assignments, etc. One of the metrics they use for assessment is unit test occurrence in the project using test ratio (number of source lines of code in test files to the number of source lines of code in all source files) to quantify the extent of the unit testing effort. Package imports of \textit{JUnit} and \textit{TestNG} frameworks were searched to identify tests in the project. This method could be useful when looking for the occurrence of specific testing frameworks in the code.


\section{Method}
\label{sec:method}
First of all, it is necessary to find suitable projects containing test cases. Thus, metadata of all GitHub open-source projects was obtained via GHTorrent~\cite{Gousi13} (section \ref{sub:data-source}) due to their high availability. GHTorrent collects projects' metadata from GitHub, one of the biggest project-sharing platform in the world. The experiment was limited to projects with Java as the primary language. Searching for testing frameworks' imports~\cite{10.1145/3030207.3030226} or files containing the word ``test'' in the filename~\cite{10.1145/3030207.3030213} are common test class identification techniques. 

Because our main goal for the future is to improve production code comprehension from a particular test case, we go deeper in this study and try to identify specific test cases (not only test classes), therefore, it is necessary to consider whether the searching for the word ``test'' is appropriate. Keep in mind, that the aim is not to count the number of test cases in a project. Otherwise, we could run tests via an automated build tool (e.g. ant, maven, or gradle) and collect the number of tests. In that case, the issue is that building such open-source projects often fail~\cite{sulir2020large} and we need to build every single project and run tests what is a time-consuming task. In this paper, we try to count and especially find the location of such test cases.

Since the testing process can also be denoted by other keywords (e.g. verify\footnote{See Mockito \texttt{verify()} method used for soft assertions: \url{https://javadoc.io/static/org.mockito/mockito-core/3.11.2/org/mockito/verification/VerificationMode.html}}, examine, etc.), an in-depth analysis (section \ref{sec:test-denotation}) of testing process denotation in various foreign languages was performed, which showed that searching for the word ``test'' is suitable. Due to the limitations of the GitHub Search API, it was possible to search only one word across all Github Java projects.

As the framework is assumed to influence developer thinking and test case implementation, a list of 50 unit testing frameworks for Java (section \ref{sub:frameworks}) has been created. Because the goal is to detect customized testing practices compared with framework--based ones in existing projects, it is not possible to use an automated method, and since it is not possible to manually analyze all GitHub projects, we need to select the most suitable ones. Based on the meaning of the word ``test'' we assume that there will be a correlation between the number of occurrences of the word ``test'' (in file content or filename) and the number of test cases. Therefore, three datasets were created using the searching GitHub API for (section \ref{sub:gathering}):
\begin{enumerate}
	\item the word ``test'' in filename,
    \item the word ``test'' in file content,
    \item frameworks' imports in file content (38 frameworks). 
\end{enumerate}

Every single project was searched as mentioned above, 4.3 million projects in total. It is possible to expect that the more occurrences of the word ``test'' in the project, the more test cases will be present in it and the more we will learn from it in the future. Therefore, projects with the highest occurrence of the word ``test'' (in file content or filename) or with the highest occurrence of a specific framework's import were selected for manual analysis. By searching for ``test'' regardless of the framework, we were also able to analyze testing practices without using any third-party framework. Because GitHub contains many projects that are not relevant, e.g. testing, homework, or cloned projects, rules for searching relevant projects have been defined (section \ref{sub:relevant-projects}), resulting in a set of projects used for manual and automated analysis. A script for automated analysis was created to partially automate the identification of test cases (see section \ref{sub:repo-analysis}). All methodology details are described in the following sections.

\subsection{Data Source}
\label{sub:data-source}
To provide conclusions that are as general as possible, it would be ideal to analyze all types of projects, i.e. proprietary and open source. Because of limited access to proprietary projects, this experiment is focused exclusively on open source projects. GitHub\footnote{https://github.com/} has become one of the most popular web-based services to host both proprietary and mostly open-source projects, therefore, we can consider it a suitable source of projects. It provides an open Application Programming Interface (API)\footnote{https://docs.github.com/en/rest} allowing one to work with all public projects (with small exceptions).

To avoid the latency of the official API, the GitHub Archive project\footnote{https://www.gharchive.org/} stores public events from the GitHub timeline and publishes them via Google BigQuery. Downloading via Google BigQuery is charged, therefore, \textit{GHTorrent}~\cite{Gousi13} was used instead, which provides a mirror of GitHub projects' metadata. It monitors the GitHub public event timeline, retrieves contents and dependencies of every event, and requests GitHub API to store project data into the database. That includes general info about projects, commits, comments, users, etc. The study data mining started in May 2019, therefore, the last MySQL dump\footnote{https://ghtorrent.org/downloads.html} \texttt{mysql-2019-05-01} has been used.

\subsection{Denotation of the Word ``test''}
\label{sec:test-denotation}
Leitner et al.~\cite{10.1145/3030207.3030213} searched for tests only in \texttt{src/test} directory and test classes identified manually. However, the tests can be placed in any project’s directory (e.g. Android\footnote{https://developer.android.com/} uses \texttt{src/androidTest}). Another approach is to search for \textit{``test''} string in filenames as executed by Kochhar et al.~\cite{7102609} because they assumed that the tests would be exclusively in files containing the case-insensitive \textit{``test''} string. As in the previous case, best practices lead the developer to use ``test'' in the file name, but it is not mandatory. For this reason, the most accurate should be searching for the word ``test'' in the file content. Of course, firstly it is necessary to consider whether the word ``test'' is the right one for searching. Therefore, the meaning of the word ``test'' using Google Translate\footnote{https://translate.google.com/} was verified in 109 different languages (all available by Google) as follows:
\begin{description}
\item [\textbf{1. From English to foreign language and back to English}]~\\
Using this method the most frequent\footnote{Frequency determined by Google Translate service, indicates how often a translation appears in public documents: 3 - high; 2 - middle; 1 - low frequency.} meanings of the word ``test'' in a foreign language were obtained. By translating them back to English we found out which foreign language translations correspond to the original word ``test''.
\item [\textbf{2. From foreign language to English and back to foreign language}]~\\
The opposite approach was used to find whether the string ``test'' has a meaning in a particular foreign language. The word was translated into English and all its meanings were verified against the available translation alternatives in the given language.
\end{description}

\begin{figure}
  \includegraphics[width=\textwidth]{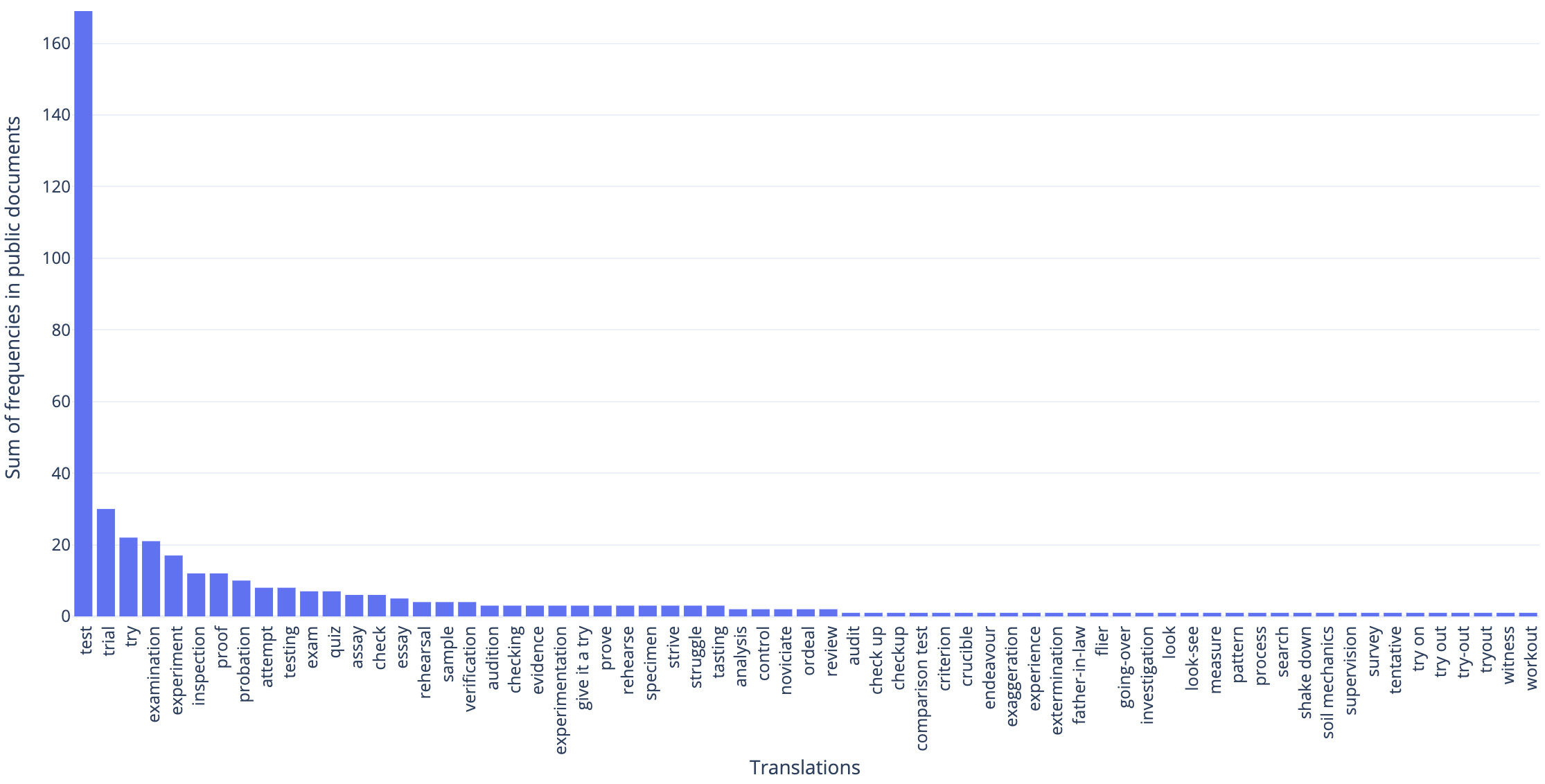}
\caption{Sum of reverse translation frequency of the word ``test'' in public documents of different languages.}
\label{fig:translations}       
\end{figure}

Multiple translations ensured that the correct meaning of the word in a particular language was understood. Using the \nth{1} method it was found out that word sets related to the testing process of different foreign languages are mostly translated as ``test'' in English, see Figure \ref{fig:translations}. This means that when a foreign developer would like to express something related to testing (e.g. to write a test case), he/she will use mostly the word ``test''. In this meaning, it is the first choice when searching test cases by a string. Occasionally occurred meaning outside of testing area, e.g., \textit{essay}, \textit{audition} or \textit{flier}. Because such meanings occurred only infrequently, they can be omitted. There were also 14 languages that did not include the word ``test'' in their reverse translation at all, but its meaning was rather denoting \textit{examination}, \textit{check} or \textit{quiz}.

A total of 44 languages used non-Latin charset. For these languages, the \nth{2} approach did not make sense to use. For the remaining languages, the meaning was completely identical in 43 languages and the same or similar meaning in 20 cases. We found only 2 languages (Hungarian\footnote{\url{https://translate.google.com/?sl=hu&tl=en&text=test}} and Latvian\footnote{\url{https://translate.google.com/?sl=lv&tl=en&text=test}}), in which the word ``test'' has a completely different meaning, such as \textit{body}, \textit{hew}, or \textit{tool} (nothing related to testing). The analysis shows that the word ``test'' will refer to the testing process in the code and the meaning can vary in very rare cases. Only the word ``test'' will be searched for in this study because of the rate limitations of the GitHub API (explained in section \ref{sub:gathering}).

\subsection{Java Testing Frameworks}
\label{sub:frameworks}
The crucial question is whether developers are motivated to use the word ``test'' in their code. The developer is often influenced by a testing framework, which leads him or her to different habits. As a part of this study, we analyzed 50 Java unit testing frameworks, extensions, and support libraries (see Table \ref{tab:test-frameworks}) to determine whether the use of the word ``test'' during test implementation is optional, recommended, or mandatory. The list was created from different sources, such as blogs, technical reports, research papers, etc. 

\begin{table}
\caption{Analyzed unit testing frameworks and extensions for Java.}
\label{tab:test-frameworks}       
\scalebox{0.65}{
\begin{tabular}{llcccc}
\hline\noalign{\smallskip}
\textbf{Name} & \textbf{\begin{tabular}[l]{@{}l@{}}Package for import\end{tabular}} & \textbf{\begin{tabular}[c]{@{}c@{}}Framework\\ type\end{tabular}} & \textbf{\begin{tabular}[c]{@{}c@{}}First\\ version\end{tabular}} & \textbf{\begin{tabular}[c]{@{}c@{}}Last\\ commit\end{tabular}} & \textbf{\begin{tabular}[c]{@{}c@{}}Must include\\ "test"\end{tabular}} \\
\noalign{\smallskip}\hline\noalign{\smallskip}
SpryTest & N/A & U & N/A & \begin{tabular}[c]{@{}c@{}}N/A\\ (archived)\end{tabular} & N/A \\
Instinct & N/A & B & 24.01.2007 & \begin{tabular}[c]{@{}c@{}}07.03.2010\\ (archived)\end{tabular} & N/A \\
\begin{tabular}[c]{@{}l@{}}Java Server-Side\\ Testing framework\\ (JSST)\end{tabular} & N/A & U & 17.11.2010 & \begin{tabular}[c]{@{}c@{}}17.11.2010\\ (archived)\end{tabular} & $\blacksquare$ \\
NUTester & N/A & U & 05.02.2009 & \begin{tabular}[c]{@{}c@{}}27.03.2012\\ (archived)\end{tabular} & N/A \\
SureAssert & N/A & A & 29.05.2011 & \begin{tabular}[c]{@{}c@{}}04.02.2019\\ (archived)\end{tabular} & N/A \\
Tacinga & N/A & U & 14.02.2018 & \begin{tabular}[c]{@{}c@{}}22.02.2018\\ (archived)\end{tabular} & N/A \\
Unitils & N/A & U & \begin{tabular}[c]{@{}c@{}}29.09.2011\\ (v3.2)\end{tabular} & \begin{tabular}[c]{@{}c@{}}08.10.2015\\ (archived)\end{tabular} & N/A \\
Cactus & org.apache.cactus & U & 11.2008 & \begin{tabular}[c]{@{}c@{}}05.08.2011\\ (archived)\end{tabular} & $\blacksquare$ \\
Concutest & N/A & U & 30.04.2009 & \begin{tabular}[c]{@{}c@{}}12.01.2010\\ (archived)\end{tabular} & $\blacksquare$ \\
Jtest & N/A & G & 1997 & \begin{tabular}[c]{@{}c@{}}21.05.2019 \\ (last release)\end{tabular} & $\blacksquare$ \\
Randoop & N/A & G & 23.08.2010 & 05.05.2020 & $\blacksquare$ \\
EvoSuite & N/A & G & \begin{tabular}[c]{@{}c@{}}25.12.2015\\ (v1.0.2)\end{tabular} & 30.04.2020 & $\blacksquare$ \\
JWalk & N/A & G & 19.05.2006 & 14.06.2017 & $\blacksquare$ \\
TestNG & org.testng & U & \begin{tabular}[c]{@{}c@{}}31.07.2010\\ (v5.13)\end{tabular} & 11.04.2020 & $\blacksquare$ \\
Artos & com.artos & U & 22.09.2018 & 19.04.2020 & $\blacksquare$ \\
JUnit 5 & org.junit & U & 10.09.2017 & 02.05.2020 & $\blacksquare$ \\
JUnit 4 & org.junit & U & 16.02.2006 & 10.04.2020 & $\blacksquare$ \\
JUnit 3 & junit.framework & U & N/A & N/A & $\blacksquare$ \\
BeanTest & info.novatec.bean-test & U & 23.04.2014 & 02.05.2015 & $\blacksquare$ \\
GrandTestAuto & org.GrandTestAuto & U & 21.11.2009 & 22.01.2014 & $\blacksquare$ \\
Arquillian & org.jboss.arquillian & U & 10.04.2012 & 21.04.2020 & $\blacksquare$ \\
EtlUnit & \begin{tabular}[l]{@{}l@{}}org.bitbucket\\.bradleysmithllc.etlunit\end{tabular} & U & \begin{tabular}[c]{@{}c@{}}02.12.2013\\ (v2.0.25)\end{tabular} & 04.04.2014 & $\blacksquare$ \\
HavaRunner & com.github.havarunner & U & 16.12.2013 & 08.06.2017 & $\blacksquare$ \\
JExample & ch.unibe.jexample & U & 2008 & N/A & $\blacksquare$ \\
Cuppa & org.forgerock.cuppa & U & 22.03.2016 & 01.10.2019 & $\blacksquare$ \\
DbUnit & org.dbunit & U & 27.02.2002 & 24.02.2020 & $\blacksquare$ \\
GroboUtils & net.sourceforge.groboutils & U & 20.12.2002 & 05.11.2004 & $\blacksquare$ \\
JUnitEE & org.junitee & U & \begin{tabular}[c]{@{}c@{}}23.07.2001\\ (v1.2)\end{tabular} & 11.12.2004 & $\blacksquare$ \\
Needle & de.akquinet.jbosscc.needle & U & N/A & 16.11.2016 & $\blacksquare$ \\
OpenPojo & com.openpojo & U & 13.10.2010 & 20.03.2020 & $\blacksquare$ \\
Jukito & org.jukito & U/M & 25.01.2011 & 17.04.2017 & $\blacksquare$ \\
Spring testing & org.springframework.test & M/U & 01.10.2002 & 06.05.2020 & $\blacksquare$ \\
Concordion & org.concordion & U/SbE & \begin{tabular}[c]{@{}c@{}}23.11.2014\\ (v1.4.4)\end{tabular} & 27.04.2020 & $\square$ \\
Jnario & org.jnario & B & 23.07.2014 &  & $\square$ \\
Cucumber-JVM & io.cucumber & B & 27.03.2012 & 04.05.2020 & $\square$ \\
Spock & spock.lang & B & 05.03.2009 & 01.05.2020 & $\square$ \\
JBehave & org.jbehave & B & 2003 & 23.04.2020 & $\square$ \\
JGiven & com.tngtech.jgiven & B & 05.04.2014 & 10.04.2020 & $\blacksquare$ \\
JDave & org.jdave & B & 18.02.2008 & 17.01.2013 & $\square$ \\
beanSpec & org.beanSpec & B & 15.09.2007 & \begin{tabular}[c]{@{}c@{}}27.06.2014\\ (alpha)\end{tabular} & $\square$ \\
EasyMock & org.easymock.EasyMock & M & 2001 & 10.04.2020 & $\blacksquare$ \\
JMock & org.jmock & M & 10.04.2007 & 23.04.2020 & $\blacksquare$ \\
JMockit & org.jmockit & M & 20.12.2012 & 13.04.2020 & $\blacksquare$ \\
Mockito & org.mockito & M & 2008 & 30.04.2020 & $\blacksquare$ \\
Mockrunner & com.mockrunner & M & 2003 & 16.03.2020 & $\blacksquare$ \\
PowerMock & org.powermock & M & \begin{tabular}[c]{@{}c@{}}28.05.2014\\ (v1.5.5)\end{tabular} & 30.03.2020 & $\blacksquare$ \\
AssertJ & org.assertj & A & 26.03.2013 & 05.05.2020 & $\blacksquare$ \\
Hamcrest & org.hamcrest & A & 01.03.2012 & 06.05.2020 & $\blacksquare$ \\
XMLUnit & org.xmlunit & A & 03.2003 & 04.05.2020 & $\blacksquare$ \\ 
\noalign{\smallskip}\hline

\multicolumn{6}{l}{} \\
\multicolumn{6}{l}{\begin{tabular}[c]{@{}l@{}}Legend: U -- unit; B -- behavioural; A -- assert; M -- mock; G -- generator; \\ SbE -- specification by example\end{tabular}} \\
\end{tabular}
}
\end{table}

Because it is sometimes difficult to find the boundary between unit and integration testing, the table lists frameworks supporting integration testing under the \textit{unit testing} category. Information about the first version and the last commit may be interesting in terms of the framework lifetime and its occurrence in projects. Projects marked as \textit{archived} or \textit{test generators} in Table \ref{tab:test-frameworks} were excluded from further analysis for the following reasons:
\begin{enumerate*}
\item archived projects usually had unavailable documentation or were never released;
\item test generators produce tests that are not based on the programmer's mental model but are generated automatically (semi-randomly), which is not interesting from the code comprehension point of view.
\end{enumerate*}

It can be seen that 37 of 50 frameworks require the word ``test'' as method/class annotation (\texttt{@Test}) or part of its name (\texttt{testMethod}, \texttt{methodTest}). The listed frameworks are mostly extensions that depend on one of the base frameworks, such as \textit{JUnit} or \textit{TestNG}. Different versions of \textit{JUnit} are listed separately because test labeling differs between them (annotations vs. method name format). A deeper analysis of frameworks' JavaDocs revealed that many frameworks include other classes, methods, or annotations that include the word ``test'' in their names. Although the use of these methods is not mandatory, it may support the search. 

\subsection{Searching Projects and Data Gathering}
\label{sub:gathering}
\begin{figure}
  \includegraphics[width=\textwidth]{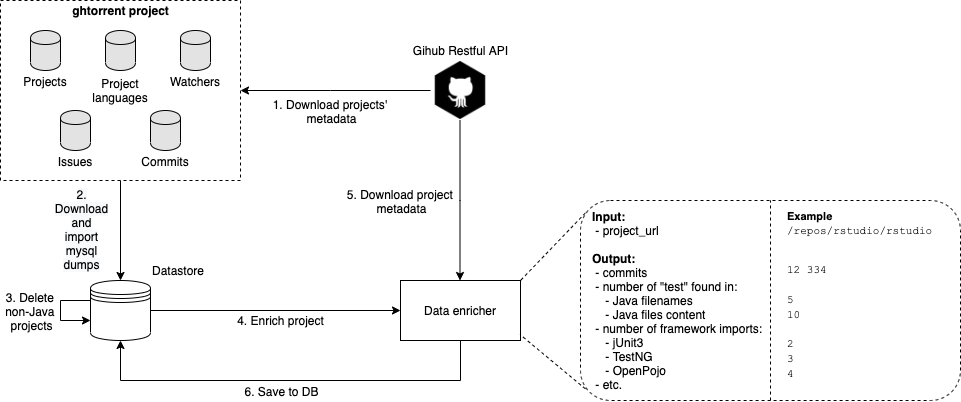}
\caption{The GitHub data mining process for the study.}
\label{fig:method}
\end{figure}

The whole process of data gathering can be seen in Figure \ref{fig:method}. GHTorrent provided 140 million GitHub projects. From this set all deleted, non-Java, or duplicated projects have been removed. After cleaning the initial data, a total of 6.7 million projects were kept for further analysis.

GHTorrent contained only basic metadata about the projects, which was not sufficient for our needs. Given the meaning of the word ``test'' (see section \ref{sec:test-denotation}) we searched for it across all projects. The GitHub API provides a code search\footnote{\url{https://docs.github.com/en/rest/reference/search}} endpoint, which index only original repositories. Repository forks are not searchable unless the fork has more stars than the parent repository. If the project has been detected as deleted, private, or blocked by GitHub during querying code search, it has been not considered. Finally, a total of 4.3 million projects were included. For each project, two requests to the GitHub code search API were called, as presented in Table \ref{tab:github-requests}. The GitHub code search API had the following limitations:
\begin{itemize}
\item up to 1,000 results for each search;
\item up to 30 requests per minute (authenticated user);
\item global requests rate limited at 5,000 requests per hour;
\item only files smaller than 384 KB and repositories with fewer than 500,000 files are searchable.
\end{itemize}

\begin{table}
\caption{The GitHub API requests used to search the string \textit{``test''} in a project.}
\label{tab:github-requests}       
\begin{tabular}{lll}
\hline\noalign{\smallskip}
Search \textit{``test''} in & Example request at \texttt{https://api.github.com/search/code} \\
\noalign{\smallskip}\hline\noalign{\smallskip}
Java files content & \texttt{\href{https://api.github.com/search/code?q=test+in:file+language:java+repo:apache/camel}{?q=test+in:file+language:java+repo:apache/camel}} \\
Java filenames & \texttt{\href{https://api.github.com/search/code?q=filename:test+language:java+repo:apache/camel}{?q=filename:test+language:java+repo:apache/camel}} \\
\noalign{\smallskip}\hline
\end{tabular}
\end{table}

\subsubsection{Code Search Strategy}
GitHub indexes only the default branch code (usually \texttt{master}), so the whole analysis was performed only using the default branch. The string ``test'' can also be a part of other words, e.g. \textit{fas\textbf{test}}, \textit{las\textbf{test}}, \textit{thisis\textbf{test}framework}. There exist 532 such words containing ``test''\footnote{\url{https://www.thefreedictionary.com/words-containing-test}} in total. To avoid inaccuracies when searching for a word of the selected string, false positives must be excluded from the search. When using regular GitHub search, the search term will appear in the results when driven by the following rules:
\begin{itemize}
\item string uses camel case convention without numbers\footnote{Numbers can be used, but they are not considered as individual words, e.g. \texttt{123Test} or \texttt{test123} will not be found.}, e.g., \texttt{myTest},
\item string uses snake case convention,  e.g., \texttt{my\_test}, \texttt{test\_123};
\item string includes a delimiter or special character (space, ., \#, \$, @, etc.), e.g., \texttt{test.delimiter}, \texttt{@Test};
\item search is case insensitive, e.g. \texttt{Test sentence}, \texttt{test sentence}.
\end{itemize}

GitHub considers as Java language file any file with \texttt{.java} or \texttt{.properties} extensions. The same search rules apply to both search types: file content and filename search. Obviously, according to the above rules, GitHub search automatically filters the results, therefore, unwanted words containing the string ``test'' do not appear in the results, but neither the words \texttt{testing} or \texttt{testsAllMethods} will be matched. 

\subsubsection{Selection of Relevant Projects}
\label{sub:relevant-projects}
When searching for different testing types, the effort is to go through as many projects as possible. Because GitHub contains millions of repositories, it is a challenge to choose the projects that can be the most instructive and filter out irrelevant ones. To make the selection as objective as possible, we planned to use \textit{reaper} tool~\cite{Munaiah2017}, which can assess a GitHub repository in collaboration with \textit{GHTorrent} using project metadata and code: architecture, community, continuous integration, documentation, history, issues, license, and unit testing. By evaluating all these metrics (see~\cite{Munaiah2017} for details), reaper tags a particular repository as a real software project and thus exclude example projects, forks, irrelevant ones, etc.

Many assessment attributes of the \textit{reaper} tool\footnote{\url{https://github.com/RepoReapers/reaper}} require project files to be available, so each project needs to be cloned or downloaded as an archive. For large projects, it can be gigabytes of data and the size of the project subsequently affects the length of the analysis. To find out whether \textit{reaper} will be beneficial for our study, a manual analysis of 50 projects was performed and the results were compared with the evaluation by \textit{reaper}. All available evaluation attributes were selected except for unit tests assessment because it was limited to \textit{JUnit} and \textit{TestNG} frameworks. The thresholds and weights of particular attributes defined by the developers of the tool were preserved because these values were considered empirically confirmed.

Because we want to select a sample of projects from which we would learn the most, projects with the highest number of files containing the word ``test'' in its body and filename were selected for the comparison. The same attributes as used by the \textit{reaper} were taken into account in the manual evaluation, but the relevance of the project for this study was assessed by an observer. Evaluation of 50 projects using the \textit{reaper} tool took 10 days, with the most time being spent on evaluating the project architecture. Many repositories with the highest ``test'' presence in file content or filename were actually identified as \textit{Subversion} (SVN) mirrors\footnote{e.g.
\url{https://github.com/zg/jdk}, \url{https://github.com/dmatej/Glassfish}, \url{https://github.com/svn2github/cytoscape}} by manual analysis and because there were multiple copies of the same code (caused by the SVN's branching style), the projects were not relevant, but the \textit{reaper} assessed such projects as suitable. According to this significant issue, important projects could be lost by assessing project in an automated manner, so it was concluded that it is more efficient to select projects manually driven by the following rules, inspired by existing research:
\begin{itemize}
    \item \textbf{Priority} was given to projects with the highest number of the word ``test'' in the project (in file content and filename). According to~\cite{6606557} we can expect the presence of tests in popular projects. If it is assumed that the word ``test'' will be correlated with the number of test cases in the project, large and long maintained projects are expected, which authors consider the best sample for the study.
    \item \textbf{History}, as evidence of sustained evolution. Projects under 50 commits were excluded (inspired by the \textit{reaper}) because they represented small or irrelevant projects. Those projects that contained a large number of commits (more than 1 000 per day), considered committed by a robot, were also excluded.
    \item \textbf{Originality} was evaluated by comparing the \texttt{readme} file for similarities in other repositories. By such comparison, it is possible to detect clones and similar repositories~\cite{7884605}. Jiang et al.~\cite{Jiang2017} found that developers clone repositories to submit pull requests, fix bugs, add new features, etc. The problem is that developers often do not create forks but project clones (a manual copy of a project), but \texttt{readme} file is often unchanged. 
    \item \textbf{Community}, as evidence of collaboration, was assessed by the number of contributors in the project. The more developers participate in the project, the more likely it is that the (testing) code will be written in a different style.
\end{itemize}

\subsubsection{Searching Java Testing Frameworks}
We were inspired by the work of Stefan et al.~\cite{10.1145/3030207.3030226}, who searched for Java performance testing frameworks imports to assess performance testing practices. In our work we are interested in the impact of testing frameworks on test writing, so we also searched for imports of all testing frameworks in Table \ref{tab:test-frameworks} (excluding generators and archived projects).

Using the search for imports we obtained projects with different testing frameworks. Only projects that contained the word ``test'' in the Java file body at least once were queried. Because there was a large number of requests (37 per single project), the project set was limited to 500,000, ordered by the number of Java files containing the word ``test'' in its body, using the following request: 

\mint[fontsize=\footnotesize]{bash}|https://api.github.com/search/code?q="org.testng"+in:file+language:java+repo:apache/camel|

For each testing framework, we created a separate list of projects, sorted by the occurrence of the word ``test'' in the project, to find projects with a high number of test cases if possible. Original repositories of the searched framework were removed from the analysis (e.g. when searching for JUnit, the original JUnit framework repository was excluded). Subsequently, the selection of relevant projects was performed according to the steps mentioned in the section \ref{sub:relevant-projects}. For some frameworks, e.g. \textit{JExample}\footnote{\url{https://github.com/akuhn/jexample}}, which were created as a part of the research~\cite{10.1007/978-3-540-68255-4_8}, no software repositories with business focus were found and as a consequence, it was necessary to include also example, homework, or cloned/forked ones, if the original one was not publicly available.

\subsection{Repository Analysis}
\label{sub:repo-analysis}
Three different data sets were received by searching via GitHub API:
\begin{enumerate*}
    \item the word ``test'' in filename,
    \item the word ``test'' in file content,
    \item frameworks' imports in file content. 
\end{enumerate*}
The first four relevant and top projects (highest ``test'' or framework's import string occurrence) were manually investigated from each set to find out the test writing practices. The projects were cloned\footnote{\texttt{git clone}} and to keep the consistency between the ``test'' search and the manual analysis, the project was reverted to the timestamp of GitHub API download using the following command:

\mint[fontsize=\footnotesize]{bash}|git checkout `git rev-list -n 1 --before="<DOWNLOADED_AT>" "<DEFAULT_BRANCH>"`|

For each project, all files with the word ``test'' in content or filename, or framework's import in file content has been selected as possible option for manual analysis. The project files that contained the largest occurrence of the word ``test'' and framework's import in their content (expected a higher number of tests) were analyzed first. During the investigation of tests from different authors and projects, we created an automated supportive method for detecting the number of test cases in a file. It does not require compiling the code, such as for computing code coverage, or building abstract syntax tree (AST), e.g. indexing in an IDE.

Regardless of the framework, it is advisable to investigate the count of the following attributes of a source file containing the word ``test'':
\begin{enumerate}
\item \textit{Annotations \texttt{@Test}} --- very popular mostly thanks to \textit{JUnit} and \textit{TestNG}.
\item \textit{Methods containing \texttt{test} in the beginning of the name} --- best practices leads developers to use this convention (also for historical purposes).
\item \textit{Methods containing \texttt{Test} in the end of the name} --- an alternative of previous one.
\item \textit{Public methods} --- possibly all public methods of a test class can be considered as tests.
\item \textit{Occurrence of \texttt{main}} --- customized testing solutions are executed via \texttt{main()}.
\item \textit{File path containing \texttt{test}} --- should relate to testing.
\item \textit{Classes containing \texttt{\$} in the name} --- the character \texttt{\$} in a class name mostly denotes a generated code\footnote{\url{https://docs.oracle.com/javase/specs/jls/se11/html/jls-3.html\#jls-3.8}} that should not be analyzed.
\item \textit{Total number of \texttt{test} occurrence in file content} --- to reveal the relation between executable test cases and the word ``test'' presence in the content.
\end{enumerate}

All listed metrics (counts of occurrence in a file) were saved for each analyzed file. The pseudocode for collecting mentioned metrics can be seen in Listing \ref{lst:alogrithm} (implementation available at GitHub\footnote{\url{https://github.com/madeja/unit-testing-practices-in-java/blob/master/AnalyzeProjectCommand.php}}). The presented algorithm is partly the result of the study because it was created in parallel with the manual analysis. The manual analysis complements the algorithm implementation and vice versa. This algorithm was used to evaluate the test identification for each Java file containing the word ``test''. Subsequently, the automated identification was checked during the manual analysis to determine the correct number of test cases and the metric used for the calculation (e.g., the number of annotations and public methods can be the same, but the relevant number of tests can only come from one of them). It is necessary to identify the number of particular test cases to link a specific test case with the unit under test (UUT) and its specific method. Each test case is likely to represent a unique use case and thus unique information to enrich the production code.

\begin{listing}[t]
\begin{minted}{text}
Algorithm predictTests(filePath)
    Input: File path to analyze.
    Output: List of statistical data
  
    content := load filePath content and remove comments
    nonClassContent := remove all class content, keep only content outside of it
        such as imports or class annotations
    classContent := remove all content outside of the class block and keep only
        first-level methods without body using /\{([^\{\}]++|(?R))*\}/
  
    annotations := matches count of regex /@Test/ in classContent
    startsWithTest := matches count of regex 
        /public +.*void *.* +[Tt]est[a-zA-Z\\d$\_]* *\(/
        in classContent
    endsWithTest := matches count of regex 
        /public +.*void *.* +[a-zA-Z$\_]{1}[a-zA-Z\\d$\_]*Test *\(/
        in classContent
    publicMethods := matches count of regex /public +.*void +.*\(/
        in classContent
    includesMain := matches count of /public +static +void +main.*\(/
        in classContent

    hasDollar := if $ in filename, then true, else false
    testInPath := if "/test" in filePath, then true, else false
            
    if TestNG import found in content, then
        if @Test found in nonClassContent, then
            testCaseCount := publicMethods
        else
            testCaseCount := annotations
    else if JUnit4 import found in content, then
        testCaseCount := annotations
    else if JUnit3 import found in content, then
        testCaseCount := startsWithTest
    else if startsWithTest > 0, then
        testCaseCount := startsWithTest
    else if annotations > 0, then
        testCaseCount := annotations
    else 
        testCaseCount := 0
        
    return annotations, startsWithTest, endsWithTest, publicMethods
            includesMain, hasDollar, testInPath, testCaseCount
\end{minted}
\caption{Pseudocode of the algorithm for gathering metadata and identified number of tests in a Java source file.}
\label{lst:alogrithm}
\end{listing}

Gathered metadata about test case identification were analyzed from different perspectives. Test classes with the highest number of the following attributes were analyzed:
\begin{enumerate*}
\item \texttt{@Test} annotations,
\item public methods with names starting with \texttt{test},
\item public methods with names ending with \texttt{Test},
\item \texttt{main} method,
\item word ``test'' occurrence.
\end{enumerate*}
For framework--dependent searches there was an additional analysis of files with the highest framework import occurrence in the content.

\subsection{Hypothesis and Research Question Evaluation}
\label{sec:hyp-confirmation}
Our null and alternative hypotheses are:

\hypref{h:hACorrelation}\textsubscript{null}: \hACorrelationNull

\hypref{h:hACorrelation}\textsubscript{alt}: \hACorrelation

The method of calculating standard Pearson's correlation coefficient~\cite{KirchPearson} was used to confirm or reject \hypref{h:hACorrelation}. The correlation coefficient was calculated as follows:

\begin{equation}
\label{eq:pearonsr}
r = \frac{\sum (x-m_{x})(y - m_{y})}{\sqrt{\sum (x - m_{x})^{2} \sum (y - m_{y})^{2}}}
\end{equation}
where $m_{x}$ is the mean of the vector $x$ (number of ``test'' occurrences in file) and $m_{y}$ is the mean of the vector $y$ (number of identified test cases in file). We will consider the \hypref{h:hACorrelation}\textsubscript{null} as accepted when $r \in (-0.8, 0.8)$, as only absolute correlation higher than 0.8 is commonly considered significant.

To address \thref{rq:rqAFrameworks}, a class/file will be considered a customized testing solution if the following conditions are met:
\begin{itemize}
    \item must include actual tests of production code.
    \item there is at least one occurrence of the word ``test'',
    \item there is no framework import from Table~\ref{tab:test-frameworks},
    \item file contains \texttt{main()} function,
\end{itemize}

The conditions are based on Section~\ref{sub:customized} which shows that customized testing solutions were mostly implemented as common java programs using \texttt{main()} function without using any \nth{3} party framework import.


\section{Results}
\label{sec:results}

Using the automated script all repositories' files from Table \ref{tab:investigated} were processed, 38 repositories and 170,076 classes altogether, from which 803 classes and 20,340 test methods were manually investigated. Some special practices in terms of the structure of the testing code or the developer's reasoning were observed. The first 4 projects from Table \ref{tab:investigated} represent repositories with the largest occurrences of the word ``test'' in the filename, another 4 in file content and other repositories represent the top import occurrence of a particular framework. The whole dataset of searching ``test'' via GitHub API can be found at Zenodo\footnote{\url{https://doi.org/10.5281/zenodo.4566198}}.

\begin{table}
\caption{Statistics of the investigated repositories.}
\label{tab:investigated}       
\scalebox{0.71}{
\begin{tabular}{llccccccc}

\multirow{2}{*}{\textbf{Repository}} & \multirow{2}{*}{\textbf{Framework}} & \multicolumn{2}{c}{\textbf{\begin{tabular}[c]{@{}c@{}}Analyzed \\ classes\end{tabular}}} & \multicolumn{2}{c}{\textbf{\begin{tabular}[c]{@{}c@{}}Analyzed \\ tests\end{tabular}}} & \multicolumn{1}{c}{\multirow{2}{*}{\textbf{\begin{tabular}[c]{@{}c@{}}Java \\ KLOC\end{tabular}}}} & \multicolumn{1}{c}{\multirow{2}{*}{\textbf{$T_A$}}} \\
 &  & \multicolumn{1}{c}{\textbf{A}} & \multicolumn{1}{c}{\textbf{M}} & \multicolumn{1}{c}{\textbf{A}} & \multicolumn{1}{c}{\textbf{M}} & \multicolumn{1}{c}{} & \multicolumn{1}{c}{} \\
\hline\noalign{\smallskip}
openjdk/client & testng, junit & 30410 & 130 & 30410 & 1661 & 5149 & 20798 \\
SpoonLabs/astor & junit & 30331 & 36 & 30331 & 1548 & 2338 & 13324 \\
apache/camel & junit & 10438 & 81 & 10438 & 625 & 1240 & 6847 \\
apache/netbeans & testng, junit & 13056 & 78 & 13056 & 1627 & 5009 & 11908 \\
\hline
JetBrains/intellij-community & testng, junit & 20375 & 49 & 20375 & 4805 & 3842 & 13630 \\
SpoonLabs/astor & testng, junit & 30331 & 44 & 30331 & 5883 & 2338 & 13324 \\
corretto/corretto-8 & testng, junit & 13688 & 10 & 13688 & 1659 & 3638 & 10792 \\
aws/aws-sdk-java & junit & 28574 & 18 & 28574 & 302 & 3680 & 20528 \\
\hline
wildfly/wildfly & arquillian & 5109 & 24 & 5109 & 123 & 548 & 3553 \\
eclipse-ee4j/cdi-tck & arquillian & 4758 & 30 & 4758 & 139 & 97 & 2748 \\
resteasy/Resteasy & arquillian & 2821 & 13 & 2821 & 144 & 220 & 1675 \\
keycloak/keycloak & arquillian & 1681 & 16 & 1681 & 104 & 396 & 1286 \\
jsfunit/jsfunit & cactus & 222 & 13 & 222 & 125 & 21 & 142 \\
bleathem/mojarra & cactus & 737 & 16 & 737 & 250 & 171 & 556 \\
\begin{tabular}[l]{@{}l@{}}topcoder-platform\\~~~~/tc-website-master\end{tabular} & cactus & 1635 & 8 & 1635 & 42 & 366 & 1199 \\
apache/hadoop-hdfs & cactus & 325 & 4 & 325 & 20 & 101 & 282 \\
zanata/zanata-platform & dbunit & 770 & 21 & 770 & 171 & 197 & 554 \\
B3Partners/brmo & dbunit & 145 & 18 & 145 & 37 & 47 & 106 \\
gilbertoca/construtor & dbunit & 145 & 18 & 145 & 64 & 24 & 53 \\
sculptor/sculptor & dbunit & 153 & 11 & 153 & 101 & 26 & 103 \\
geotools/geotools & groboutils & 3424 & 5 & 3424 & 5 & 1272 & 3659 \\
notoriousre-i-d/ce-packager & groboutils & 107 & 11 & 107 & 75 & 46 & 91 \\
tliron/prudence & groboutils & 16 & 2 & 16 & 3 & 13 & 11 \\
MichaelKohler/P2 & jexample & 36 & 12 & 36 & 53 & 4 & 24 \\
akuhn/codemap & jexample & 132 & 15 & 132 & 286 & 41 & 112 \\
wprogLK/TowerDefenceANTS & jexample & 17 & 3 & 17 & 50 & 9 & 12 \\
rbhamra/Jboss-Files & needle & 44 & 21 & 44 & 30 & 5 & 30 \\
akquinet/mobile-blog & needle & 19 & 10 & 19 & 33 & 2 & 10 \\
s-case/s-case & needle & 46 & 15 & 46 & 13 & 39 & 33 \\
dbarton-uk/population-pie & needle & 7 & 6 & 7 & 16 & 1 & 4 \\
abarhub/rss & openpojo & 26 & 2 & 26 & 3 & 6 & 20 \\
\begin{tabular}[l]{@{}l@{}}BRUCELLA2\\~~~~/Prescriptions-Scolaires\end{tabular} & openpojo & 25 & 19 & 25 & 40 & 10 & 18 \\
jpmorganchase/tessera & openpojo & 382 & 8 & 382 & 12 & 45 & 234 \\
tensorics/tensorics-core & openpojo & 161 & 3 & 161 & 1 & 24 & 85 \\
\begin{tabular}[l]{@{}l@{}}orange-cloudfoundry\\~~~~/static-creds-broker\end{tabular} & jgiven & 21 & 11 & 21 & 33 & 2 & 16 \\
eclipse/sw360 & jgiven & 175 & 4 & 175 & 51 & 56 & 161 \\
\begin{tabular}[l]{@{}l@{}}Orchaldir\\~~~~/FantasyWorldSimulation\end{tabular} & jgiven & 54 & 13 & 54 & 198 & 7 & 37 \\
kodokojo/docker-image-manager & jgiven & 11 & 5 & 11 & 8 & 3 & 8 \\
\hline\noalign{\smallskip}
\textbf{SUM} & \textbf{} & \textbf{170076} & \textbf{803} & \textbf{363730} & \textbf{20340} & \textbf{31033} & \textbf{127973} \\

\multicolumn{8}{l}{} \\
\multicolumn{8}{l}{\begin{tabular}[c]{@{}l@{}}Legend: A -- processed automated; M -- investigated manually; KLOC -- kilo of lines of code; \\ ~~~~~~~~~~~$T_A$ -- average time of automated test case detection in ms.\end{tabular}}

\end{tabular}}
\end{table}

\subsection{Accuracy of Automated Test Case Identification}
To evaluate the precision of the algorithm from Listing~\ref{lst:alogrithm}, results were compared to manual test identification of 20,340 test cases across all three datasets. Accuracy of 95.72\% for test cases detection was achieved by automated identification considering only test methods, i.e., 95.72\% of all test cases were correctly identified. Considering all 28,975 methods of manually analyzed files (with non-testing ones) a total accuracy of 96.97\% was achieved with the sensitivity of
\begin{equation}
Sensitivity = \frac{true~positives}{true~positives + false~negatives} = \frac{19600}{19600 + 62} = 0.9968
\end{equation}
and specificity of
\begin{equation}
Specificity = \frac{true~negatives}{true~negatives + false~positives} = \frac{8498}{8498 + 815} = 0.9125
\end{equation}
Most false positives and false negatives occurrences were caused by customized testing solutions, e.g., when tests were performed directly from the \texttt{main()} function by calling methods of the class. If the naming conventions of the called (testing) methods were not governed by the principles of frameworks (e.g., prepending method name with \textit{``test''} or using public methods), not all test cases were detected in an automated way.

\subsection{Correlation Between the Number of the Word ``test'' and the Number of Test Cases in a Class}
\label{sub:correlation}
\begin{figure}
\includegraphics[width=0.75\textwidth]{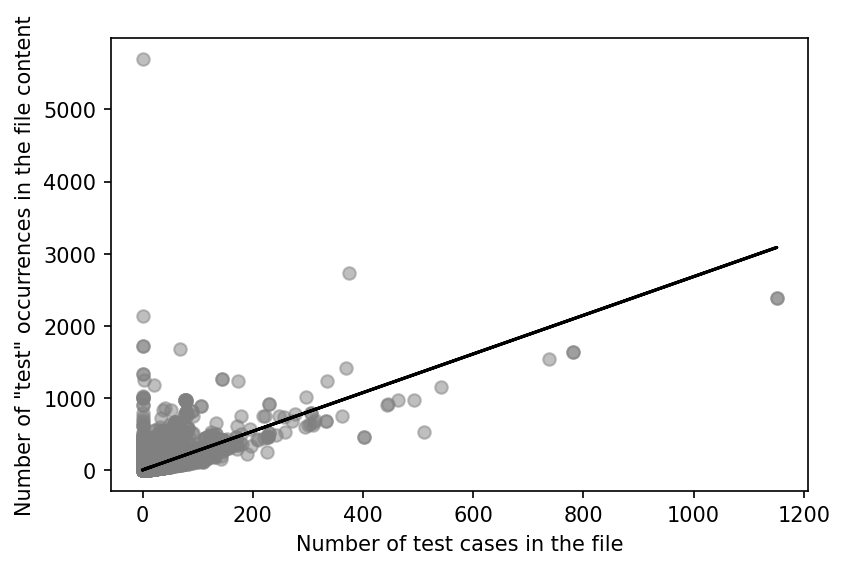}
\caption{Correlation of the word ``test'' presence and number of test cases for analyzed classes by automated script.}
\label{fig:correlation}
\end{figure}

The proposed algorithm was used to identify all tests in all Java classes of projects from Table \ref{tab:investigated}. The script was used for all Java files that contained string \textit{``test''} in the file content or the filename (in total 170,076 files). Figure~\ref{fig:correlation} shows the correlation with the linear regression line of the word ``test'' and the number of test cases in a particular class. A standard Pearson's correlation coefficient of $r = 0.655$ was reached (statistical significance $p = 0.0$, rounded on 5 decimal places), that means there is a weak correlation when considering absolute threshold $\alpha = 0.2$ defined in Section~\ref{sec:hyp-confirmation}. Nevertheless, from the perspective of finding projects containing tests, this technique is beneficial and can help future experimenters to filter projects containing tests much faster. Because projects have different numbers of test classes and use different frameworks, the detailed ratio of the word ``test'' occurrence and test case presence per project can be found at GitHub\footnote{\url{https://github.com/madeja/unit-testing-practices-in-java/blob/master/correlation-boxplot.png}}.

Due to existing research~\cite{10.1145/3030207.3030213} that identified test files using searching ``test'' in the file path, when limiting our results to files containing ``test'' in the path (120,907 files) the correlation coefficient of $r = 0.6649$ was reached. On the other hand, 49,169 classes with 3,855 test cases were discarded. Limiting results to files containing ``test'' in filename (74,530 files), we reached correlation coefficient $r = 0.7004$ with loss of 95,546 classes and 17,440 test cases. By any limitation the correlation did not significantly change, therefore, to find as many test cases as possible it is convenient to search for the word ``test'' in the file content.

Occurrence of the function \texttt{main} without the \nth{3} party testing framework (more explained in section \ref{sub:customized}) was detected in 26,205 (15.41\%) classes containing the word ``test'' in their content. The proposed algorithm in section \ref{sub:repo-analysis} successfully identified test cases in only 6\% classes of this set. Because \texttt{main} tests make up a fairly large proportion and the identification of test cases is not clear, it is necessary to investigate this testing style deeper in the future. 

\begin{mdframed}[linecolor=black,linewidth=1,nobreak=true]
\hypref{h:hACorrelation} \textit{\hACorrelation}\\

We accept \hypref{h:hACorrelation}\textsubscript{null} and reject \hypref{h:hACorrelation}\textsubscript{alt} because only weak Pearson's correlation coefficient $r=0.655$ was achieved in general. In some projects, when the correlation was calculated for each project separately, a significant correlation was achieved but so far no relationship has been found concerning the framework, the number of the word ``test'' presence in the content, or other dependencies.
\end{mdframed}

\subsection{Efficiency of the Proposed Automated Test Case Identification}
Executing a full code analysis, e.g. in an IDE, of a large project with thousands of kilo of lines of code (KLOC), is a time-consuming task. Such example is the project \texttt{openjdk/client} from Table \ref{tab:investigated}. To get faster feedback about tests in a project, the proposed algorithm was used for static source code analysis. Because the proposed automated algorithm should run as a part of an integrated development environment (IDE) extension in the future it should be fast enough. To emulate a similar environment that a developer can use, a laptop with \textit{2.3 GHz Dual-Core Intel Core i5} CPU and \textit{8 GB 2133 MHz LPDDR3 RAM} was used. In the table \ref{tab:investigated}  can be seen the average time ($T_A$) of automated analysis executed 10 times. The average time of execution was 158ms per KLOC, which authors consider as a satisfactory response time in terms of user experience for use in an IDE extension.

\subsection{Revealed Testing Practices}
In related work (section \ref{sec:bg}) there are best practices that developers can follow and therefore can be expected in the code. During the manual investigation of multiple repositories containing tests, we identified special testing practices used by developers, which are described in the following paragraphs. The listings that are given as examples come from the analyzed repositories, but the code was simplified for presentation purposes. Code listings refer to GitHub\footnote{\url{https://github.com/madeja/unit-testing-practices-in-java}} repository of this paper.

\subsubsection{Testing Using \nth{3} Party Frameworks}

\textbf{Regular test.} Tests that follow best practices and avoid test smells fall into this category. They represent the most of occurrences in the projects and since these approaches are already described in the available literature~\cite{nayyar2019instant,lewis2017software,garcia2017mastering}, this group will not be given detailed attention. However, the basic aspect of such tests is that information about context and evaluation are available directly in the particular test method (considering also test setup, teardown, and fixtures), thanks to which the test
comprehension is straightforward.

\textbf{Master test.} This testing code style represents test classes which contain only one executable test method (see GitHub\footnote{\url{https://github.com/madeja/unit-testing-practices-in-java/blob/master/examples/c_masterTest.java}}). \textit{JUnit} will consider only the \texttt{all()} method as a test case because it is annotated with \texttt{@Test} annotation. Other methods are considered auxiliary ones. The problem with such a notation is the complexity of test comprehension. If the test fails, the developer only has information that the test case titled \texttt{all} failed but does not know what the test should have verified, what data was used, etc.

According to the best practices, it should be clear from the test name what the test verifies. In this context, from a semantic point of view, it is possible to consider methods as test cases on lines 1-8 (here from L1-8). The mentioned methods are crucial in terms of failure and understanding of the test, and from the method name, it is also clear what the test verifies. Another disadvantage of these test types is the \textit{assertion roulette} test smell~\cite{van2001refactoring} because iterations of the test over the input data make it difficult to determine which data caused the test failure and whether the input data do not interfere with each other between the tests.

\textbf{Reverse proxy test.} If a separate test is written for each use case, the recommendations are met, but this does not mean that it will be easy to comprehend. Some tests call one auxiliary method in multiple tests and the result is evaluated in the auxiliary method. According to the test evaluation manner, they can be divided into:
\begin{enumerate}
\item \textit{Result evaluation via method name} (see GitHub\footnote{\url{https://github.com/madeja/unit-testing-practices-in-java/blob/master/examples/c_reverseProxyMethod.java}}).
\item \textit{Result evaluation via internal object state} (see GitHub\footnote{\url{https://github.com/madeja/unit-testing-practices-in-java/blob/master/examples/c_reverseProxyObject.java}}).
\end{enumerate}

The \nth{1} approach is much more difficult to comprehend due to the high degree of abstraction. It is not clear directly from the test method code (L6-8) what is compared during the test because the input data are loaded from a file determined by the test method name (L3). In the \texttt{JetBrains/intellij-community} project, from which the example is given, the \texttt{doTest()} method is the general one and it was necessary to investigate multiple classes to comprehend how tests are evaluated. At the same time, too generic auxiliary method can result in the \textit{general fixture} test smell.

The \nth{2} approach is similar to the previous one but uses the internal state of an object (that is initialized before a particular test during test setup) or the \texttt{enum} type with different method implementations. The problem may arise when object attribute or method input parameter change the control flow. If the same test is called with different object state or input data, the test logic does not change and therefore it is the same test. However, if the control flow changes in the test, e.g. by some variable value, it can be considered as a separate test (different flow, different test). If the same help method is called more than once, it may behave like 2 different test cases, which contradicts best practices and makes the comprehension difficult.

\textbf{Multiple test execution.} Server-side applications test different use cases, which require an action after the execution of base functionality, e.g. whether the right content is shown after main test execution (see GitHub\footnote{\url{https://github.com/madeja/unit-testing-practices-in-java/blob/master/examples/c_multipleExecution.java}}). Because of using \textit{JUnit3} in the example, every public method prepended by ``test'' is considered as test case, so \texttt{testEcho()} is executed twice; as a single test case and as a part of \texttt{testA4JRedirect()}.

\subsubsection{Customized Testing Solutions}
\label{sub:customized}
Custom testing practices are classic Java programs executable via \texttt{main()} function, whose task is to verify the functionality of the production code. Such tests are often written due to the possibility of configuring the execution via command line parameters, which allows variability of test execution. On the other hand, tests should not be so environmentally dependent that they need to be configured to such an extent. The second reason for writing such tests is that they make the code with a large number of test cases more readable. Test methods are called directly from \texttt{main()} and, if necessary, also the environment setup is performed in this function. The following ways of calling test methods and objects were observed (examples can be found at GitHub\footnote{\url{https://github.com/madeja/unit-testing-practices-in-java}}):

\begin{itemize}
\item \textit{Calling methods one by one:} all testing methods are manually called from \texttt{main()} together with parameters.
\item \textit{Calling methods according to input data:} by iterating the test data, specific tests are called based on the current data.
\item \textit{Helper function that returns an array of test cases:} the helper method returns an array of instances created from abstract classes, whereas the abstract methods (which represent test cases) are implemented during the instance creation. The \texttt{main()} contains an iteration over the array of object instances.
\item \textit{Iterating values of \texttt{enum}:} similar to the previous one, but it iterates over \texttt{enum} values. When creating the \texttt{enum}, the method of test class is implemented and the data is set. The test class has its own implementation of a method and state in each iteration.
\item \textit{Calling constructor:} in the \texttt{main} function the testing class instance is created and the tests are called from the constructor.
\end{itemize}

There is a problem of how to identify such tests using an automated way and how to determine the number of tests in such a class. The \texttt{main()} function also occurs in classic tests (e.g. to run test outside of IDE or without a build automation tool\footnote{\url{https://junit.org/junit4/faq.html}}), e.g. based on \textit{JUnit} or \textit{TestNG}. The function can also be found in modified runners of testing frameworks. To clearly distinguish the presence of a customized solution without any framework, it is possible to check the presence of the framework import --- if a class contains the \texttt{main()} function and an import together, it is a runner or regular test based on the framework, not a customized solution.

Other interesting ways of writing customized tests were also observed. For example, in the \texttt{openjdk/client} repository, there were tests for trichotomous relations for which a custom \texttt{@Test} annotation was implemented (see GitHub\footnote{\url{https://github.com/madeja/unit-testing-practices-in-java/blob/master/examples/c_main1.java}}). The annotation is used to indicate the test and, at the same time, to define the type of comparison in the method (L1, L4). Thanks to the word ``test'' usage, it is possible to detect the correct number of tests, in a similar way as for \textit{JUnit}. In this example, the impact of \nth{3} party framework on the developer's customized solution is visible. There are many tests in the repository using standardized frameworks, therefore the usage of \texttt{@Test} annotation is a logical way of defining a test case. Writing tests manually using a framework would not be as effective and would be difficult to comprehend. On the other hand, such tests in large iterations can easily give rise to the \textit{assertion roulette} test smell, which makes it difficult to identify a test failure.

While in the previous case the test was evaluated using asserts, some approaches have their own error handling. E.g. in the same repository for all \textit{ResourceBundle} classes, a helper test class \texttt{RBTestFmwk} has been implemented, which represents a custom framework and test classes inherit from it. The framework provides the processing of the \texttt{main()} function parameters, performing tests, and processing results. The test methods to be performed are defined as input parameters. The disadvantage is that when performing such tests, it is necessary to know the internal structure of the class, at least method names that need to be performed. 

In general, the following risks were observed by analyzing other \texttt{main} testing methods:
\begin{itemize}
\item \textit{Execution interruption} --- If a test fails, execution may be completely interrupted and no further tests will be performed (e.g. raised exception).
\item \textit{Failure identification} --- Because testing is often performed repeatedly over different data, it can be difficult to identify the exact cause of test failure and in some cases may require debugging the test code.
\item \textit{Dependence} --- Tests often use the same sources or data for testing and may affect the results of other tests. Also, the tests are often order-dependent and the test order randomness was not found in any repository.\end{itemize}

Occurrence of the \texttt{main()} function without any \nth{3} party testing framework was detected in 26,205 (15.41\%) classes containing the word ``test'' in their content. The proposed algorithm in Section~\ref{sub:repo-analysis} successfully identified test cases in only 6\% classes of this set. The set can contain not only testing code, but also a production one. Because such classes make up a fairly large proportion and the identification of test cases is not clear due to the high diversity of writing such tests, it is necessary to carry out an extensive study dealing solely with this issue, to find a way to precisely identify such test cases.

\begin{mdframed}[linecolor=black,linewidth=1,nobreak=true]
\thref{rq:rqAFrameworks} \textit{\rqAFrameworks}\\

A total of 15\% of classes were identified as customized testing solutions. The diversity of such tests is very high, therefore, future investigation is needed. This high incidence is probably caused by the nature of big projects with a high occurrence of the word ``test'' in file content and it is assumed the use of \nth{3} party frameworks should be more common in smaller projects.
\end{mdframed}


\section{Threats to Validity}
\label{sec:threats}

\textbf{Internal validity:} The study relied on GHTorrent databank and GitHub API search algorithm to identify relevant projects. Because only projects with Java as a primary language were selected, testing practices in projects, where Java was not a major language could have been lost. Test classes that did not use the word ``test'' to indicate a test case were also lost. Searching for test cases was based on best practices and rules of the identified frameworks, but there may still exist other ways of how to identify them. The manual classification was based on observers' experiences and identification of practices out of the generally known recommendations (best practices, test smells, etc.).

Test case detection results were compared to manual ones with an accuracy of 96.97\%. As stated, it is necessary to further investigate customized testing solutions that use regular Java programs to test the production code. The implementation of such programs is often diametrically different and it is difficult to identify test cases. Real test cases were identified by the script in 6\% of classes containing \texttt{main()} function.

\textbf{External validity:} To provide generalizable results, 20k of test cases were analyzed manually and 170k by an automated way. Also, the meaning and occurrence of the word ``test'' was analyzed for different natural languages and test frameworks. The results can be used to identify test cases in Java-based projects or projects with a different programming language with the usage of similar testing conventions. Despite the presented observations, our findings, as is usual in empirical software engineering, may not be directly generalized to other systems, particularly to commercial or to the ones implemented in other programming languages.


\section{Conclusion and Future Work}
\label{sec:concl}
This paper presented an empirical study of Java open source GitHub projects to better understand how to identify test cases and their exact location in the project without the need for deep and time-consuming dynamic code analysis. An algorithm based on searching the word ``test'' in the repository files content or filenames was proposed and, at the same time, the unusual testing practices were investigated. In total 20,340 test cases in 803 classes were investigated manually and 170k classes in an automated way. We summarise the most interesting findings from our study:
\begin{itemize}
\item There is not a strong correlation between the number of occurrences of the word ``test'' and the number of test cases in a class.
\item Searching for the word ``test'' in the file content can be used to identify projects containing tests.
\item Using static file analysis, the proposed algorithm can correctly detect 97\% of test cases.
\item Approximately 15\% of the analyzed files contains ``test'' in the content together with \texttt{main()} function whose represent regular Java programs that test the production code without using any third-party framework. The success rate of identification of such test cases is very low because of implementation diversity.
\end{itemize}

Several test writing styles were found and classified, along with code samples of the analyzed repositories. Possible code comprehension defects were also mentioned. Based on these findings the following main contributions of this paper are concluded:
\begin{itemize}
    \item Possibility of fast and automated test case identification together with the exact location in the project.
    \item Finding of correlation coefficient $r = 0.655$ between the number of occurrences of the word ``test'' and the number of test cases in a file, which allows to threshold projects or files for similar analysis.
    \item Overview of observed testing practices concerning the existence of customized testing solutions with an emphasis on places in testing code usable for mining information about the production code.
\end{itemize}

As future work, we plan to find a solution for accurate identification of test cases in customized solutions, probably based on training a machine learning model with manually labeled test cases of such testing solutions. We believe that studying testing practices can help comprehend the production code more easily. Gathered data could be used for training a machine learning model to automatically recognize tests by the structure and nature of the code. At the same time, we would like to focus on mining information from tests that could support the production source code comprehension and streamline the development process.

\section{Acknowledgement}
This work was supported by project VEGA No. 1/0762/19: Interactive pattern-driven language development.

\bibliographystyle{acm}
\bibliography{bibliography.bib}

\begin{thebibliography}{10}

\bibitem{10.1145/2786805.2786843}
{\sc Beller, M., Gousios, G., Panichella, A., and Zaidman, A.}
\newblock When, how, and why developers (do not) test in their ides.
\newblock In {\em Proceedings of the 2015 10th Joint Meeting on Foundations of
  Software Engineering\/} (New York, NY, USA, 2015), ESEC/FSE 2015, Association
  for Computing Machinery, p.~179–190.

\bibitem{BISSI201645}
{\sc Bissi, W., {Serra Seca Neto}, A.~G., and Emer, M. C. F.~P.}
\newblock The effects of test driven development on internal quality, external
  quality and productivity: A systematic review.
\newblock {\em Information and Software Technology 74\/} (2016), 45 -- 54.

\bibitem{7332450}
{\sc {Butler}, S., {Wermelinger}, M., and {Yu}, Y.}
\newblock Investigating naming convention adherence in java references.
\newblock In {\em 2015 IEEE International Conference on Software Maintenance
  and Evolution (ICSME)\/} (2015), pp.~41--50.

\bibitem{CORRITORE199961}
{\sc Corritore, C.~L., and Wiedenbeck, S.}
\newblock Mental representations of expert procedural and object-oriented
  programmers in a software maintenance task.
\newblock {\em International Journal of Human-Computer Studies 50}, 1 (1999),
  61 -- 83.

\bibitem{Cruz2019}
{\sc Cruz, L., Abreu, R., and Lo, D.}
\newblock To the attention of mobile software developers: guess what, test your
  app!
\newblock {\em Empirical Software Engineering 24}, 4 (2019), 2438--2468.

\bibitem{demeyer2002object}
{\sc Demeyer, S., Ducasse, S., and Nierstrasz, O.}
\newblock {\em Object-oriented reengineering patterns}.
\newblock Elsevier, 2002.

\bibitem{1383101}
{\sc {Ellims}, M., {Bridges}, J., and {Ince}, D.~C.}
\newblock Unit testing in practice.
\newblock In {\em 15th International Symposium on Software Reliability
  Engineering\/} (2004), pp.~3--13.

\bibitem{7592412}
{\sc {Fucci}, D., {Erdogmus}, H., {Turhan}, B., {Oivo}, M., and {Juristo}, N.}
\newblock A dissection of the test-driven development process: Does it really
  matter to test-first or to test-last?
\newblock {\em IEEE Transactions on Software Engineering 43}, 7 (2017),
  597--614.

\bibitem{garcia2017mastering}
{\sc Garcia, B.}
\newblock {\em Mastering Software Testing with JUnit 5: Comprehensive guide to
  develop high quality Java applications}.
\newblock Packt Publishing Ltd, 2017.

\bibitem{6982626}
{\sc {Gopinath}, R., {Jensen}, C., and {Groce}, A.}
\newblock Mutations: How close are they to real faults?
\newblock In {\em 2014 IEEE 25th International Symposium on Software
  Reliability Engineering\/} (2014), pp.~189--200.

\bibitem{Gousi13}
{\sc Gousios, G.}
\newblock The ghtorrent dataset and tool suite.
\newblock In {\em Proceedings of the 10th Working Conference on Mining Software
  Repositories\/} (Piscataway, NJ, USA, 2013), MSR '13, IEEE Press,
  pp.~233--236.

\bibitem{7272926}
{\sc {Hemmati}, H.}
\newblock How effective are code coverage criteria?
\newblock In {\em 2015 IEEE International Conference on Software Quality,
  Reliability and Security\/} (2015), pp.~151--156.

\bibitem{10.1145/3238147.3238183}
{\sc Hilton, M., Bell, J., and Marinov, D.}
\newblock A large-scale study of test coverage evolution.
\newblock In {\em Proceedings of the 33rd ACM/IEEE International Conference on
  Automated Software Engineering\/} (New York, NY, USA, 2018), ASE 2018,
  Association for Computing Machinery, p.~53–63.

\bibitem{Jiang2017}
{\sc Jiang, J., Lo, D., He, J., Xia, X., Kochhar, P.~S., and Zhang, L.}
\newblock Why and how developers fork what from whom in github.
\newblock {\em Empirical Software Engineering 22}, 1 (2017), 547--578.

\bibitem{10.1145/2635868.2635929}
{\sc Just, R., Jalali, D., Inozemtseva, L., Ernst, M.~D., Holmes, R., and
  Fraser, G.}
\newblock Are mutants a valid substitute for real faults in software testing?
\newblock In {\em Proceedings of the 22nd ACM SIGSOFT International Symposium
  on Foundations of Software Engineering\/} (New York, NY, USA, 2014), FSE
  2014, Association for Computing Machinery, p.~654–665.

\bibitem{KirchPearson}
{\sc Kirch, W.}, Ed.
\newblock {\em Pearson's Correlation Coefficient}.
\newblock Springer Netherlands, Dordrecht, 2008, pp.~1090--1091.

\bibitem{8031982}
{\sc {Kochhar}, P.~S., {Lo}, D., {Lawall}, J., and {Nagappan}, N.}
\newblock Code coverage and postrelease defects: A large-scale study on open
  source projects.
\newblock {\em IEEE Transactions on Reliability 66}, 4 (2017), 1213--1228.

\bibitem{7081877}
{\sc {Kochhar}, P.~S., {Thung}, F., and {Lo}, D.}
\newblock Code coverage and test suite effectiveness: Empirical study with real
  bugs in large systems.
\newblock In {\em 2015 IEEE 22nd International Conference on Software Analysis,
  Evolution, and Reengineering (SANER)\/} (2015), pp.~560--564.

\bibitem{7102609}
{\sc {Kochhar}, P.~S., {Thung}, F., {Nagappan}, N., {Zimmermann}, T., and {Lo},
  D.}
\newblock Understanding the test automation culture of app developers.
\newblock In {\em 2015 IEEE 8th International Conference on Software Testing,
  Verification and Validation (ICST)\/} (2015), pp.~1--10.

\bibitem{10.1007/978-3-540-68255-4_8}
{\sc Kuhn, A., Van~Rompaey, B., Haensenberger, L., Nierstrasz, O., Demeyer, S.,
  Gaelli, M., and Van~Leemput, K.}
\newblock Jexample: Exploiting dependencies between tests to improve defect
  localization.
\newblock In {\em Agile Processes in Software Engineering and Extreme
  Programming\/} (Berlin, Heidelberg, 2008), P.~Abrahamsson, R.~Baskerville,
  K.~Conboy, B.~Fitzgerald, L.~Morgan, and X.~Wang, Eds., Springer Berlin
  Heidelberg, pp.~73--82.

\bibitem{10.1145/3030207.3030213}
{\sc Leitner, P., and Bezemer, C.-P.}
\newblock An exploratory study of the state of practice of performance testing
  in java-based open source projects.
\newblock In {\em Proceedings of the 8th ACM/SPEC on International Conference
  on Performance Engineering\/} (New York, NY, USA, 2017), ICPE ’17,
  Association for Computing Machinery, p.~373–384.

\bibitem{lewis2017software}
{\sc Lewis, W.~E.}
\newblock {\em Software testing and continuous quality improvement}.
\newblock CRC press, 2017.

\bibitem{8094467}
{\sc {Linares-Vásquez}, M., {Bernal-Cardenas}, C., {Moran}, K., and
  {Poshyvanyk}, D.}
\newblock How do developers test android applications?
\newblock In {\em 2017 IEEE International Conference on Software Maintenance
  and Evolution (ICSME)\/} (2017), pp.~613--622.

\bibitem{madeja_et_al:OASIcs:2019:10870}
{\sc Madeja, M., and Porub{\"a}n, J.}
\newblock {Tracing Naming Semantics in Unit Tests of Popular Github Android
  Projects}.
\newblock In {\em 8th Symposium on Languages, Applications and Technologies
  (SLATE 2019)\/} (Dagstuhl, Germany, 2019), R.~Rodrigues, J.~Janousek,
  L.~Ferreira, L.~Coheur, F.~Batista, and H.~G. Oliveira, Eds., vol.~74 of {\em
  OpenAccess Series in Informatics (OASIcs)}, Schloss Dagstuhl--Leibniz-Zentrum
  fuer Informatik, pp.~3:1--3:13.

\bibitem{10.1145/356835.356841}
{\sc Mayer, R.~E.}
\newblock The psychology of how novices learn computer programming.
\newblock {\em ACM Comput. Surv. 13}, 1 (Mar. 1981), 121–141.

\bibitem{Munaiah2017}
{\sc Munaiah, N., Kroh, S., Cabrey, C., and Nagappan, M.}
\newblock Curating github for engineered software projects.
\newblock {\em Empirical Software Engineering 22}, 6 (2017), 3219--3253.

\bibitem{nayyar2019instant}
{\sc Nayyar, A.}
\newblock {\em Instant Approach to Software Testing: Principles, Applications,
  Techniques, and Practices}.
\newblock BPB Publications, 2019.

\bibitem{peruma2019distribution}
{\sc Peruma, A., Almalki, K., Newman, C.~D., Mkaouer, M.~W., Ouni, A., and
  Palomba, F.}
\newblock On the distribution of test smells in open source android
  applications: an exploratory study.
\newblock In {\em CASCON\/} (2019), pp.~193--202.

\bibitem{6606557}
{\sc {Pham}, R., {Singer}, L., {Liskin}, O., {Filho}, F.~F., and {Schneider},
  K.}
\newblock Creating a shared understanding of testing culture on a social coding
  site.
\newblock In {\em 2013 35th International Conference on Software Engineering
  (ICSE)\/} (2013), pp.~112--121.

\bibitem{7503707}
{\sc {Scalabrino}, S., {Linares-Vásquez}, M., {Poshyvanyk}, D., and {Oliveto},
  R.}
\newblock Improving code readability models with textual features.
\newblock In {\em 2016 IEEE 24th International Conference on Program
  Comprehension (ICPC)\/} (2016), pp.~1--10.

\bibitem{8529832}
{\sc {Spadini}, D., {Palomba}, F., {Zaidman}, A., {Bruntink}, M., and
  {Bacchelli}, A.}
\newblock On the relation of test smells to software code quality.
\newblock In {\em 2018 IEEE International Conference on Software Maintenance
  and Evolution (ICSME)\/} (2018), pp.~1--12.

\bibitem{10.1145/3030207.3030226}
{\sc Stefan, P., Horky, V., Bulej, L., and Tuma, P.}
\newblock Unit testing performance in java projects: Are we there yet?
\newblock In {\em Proceedings of the 8th ACM/SPEC on International Conference
  on Performance Engineering\/} (New York, NY, USA, 2017), ICPE ’17,
  Association for Computing Machinery, p.~401–412.

\bibitem{sulir2020large}
{\sc Sul{\'\i}r, M., Ba{\v{c}}{\'\i}kov{\'a}, M., Madeja, M., Chodarev, S., and
  Juh{\'a}r, J.}
\newblock Large-scale dataset of local java software build results.
\newblock {\em Data 5}, 3 (2020), 86.

\bibitem{van2001refactoring}
{\sc Van~Deursen, A., Moonen, L., Van Den~Bergh, A., and Kok, G.}
\newblock Refactoring test code.
\newblock In {\em Proceedings of the 2nd international conference on extreme
  programming and flexible processes in software engineering (XP)\/} (2001),
  pp.~92--95.

\bibitem{7884645}
{\sc {Zerouali}, A., and {Mens}, T.}
\newblock Analyzing the evolution of testing library usage in open source java
  projects.
\newblock In {\em 2017 IEEE 24th International Conference on Software Analysis,
  Evolution and Reengineering (SANER)\/} (2017), pp.~417--421.

\bibitem{7884605}
{\sc {Zhang}, Y., {Lo}, D., {Kochhar}, P.~S., {Xia}, X., {Li}, Q., and {Sun},
  J.}
\newblock Detecting similar repositories on github.
\newblock In {\em 2017 IEEE 24th International Conference on Software Analysis,
  Evolution and Reengineering (SANER)\/} (2017), pp.~13--23.

\end{thebibliography}

\bio{\includegraphics[width=0.22\textwidth]{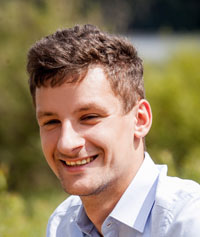}}Matej,Madeja\\ was born in 1992 in Kežmarok, Slovakia. In 2017 he graduated (MSc) at the Department of Computers and Informatics of the Faculty of Electrical Engineering and Informatics at Technical University of Košice. He defended his master’s thesis in the field of Informatics. Currently, he is a Ph.D. student in the same department. His research is focused on the improvement of program comprehension efficiency, source code testing techniques, and teaching of programming.

\bio{\includegraphics[width=0.22\textwidth]{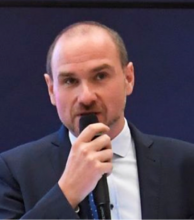}}Jaroslav,Porubän\\ is a Professor and the Head of Department of Computers and Informatics, Technical University of Košice, Slovakia. He received his MSc. in Computer Science in 2000 and his Ph.D. in Computer Science in 2004. Since 2003 he is a member of the Department of Computers and Informatics at Technical University of Košice. Currently, the main subject of his research is the computer language engineering concentrating on the design and implementation of domain-specific languages and computer language composition and evolution.

\bio{\includegraphics[width=0.22\textwidth]{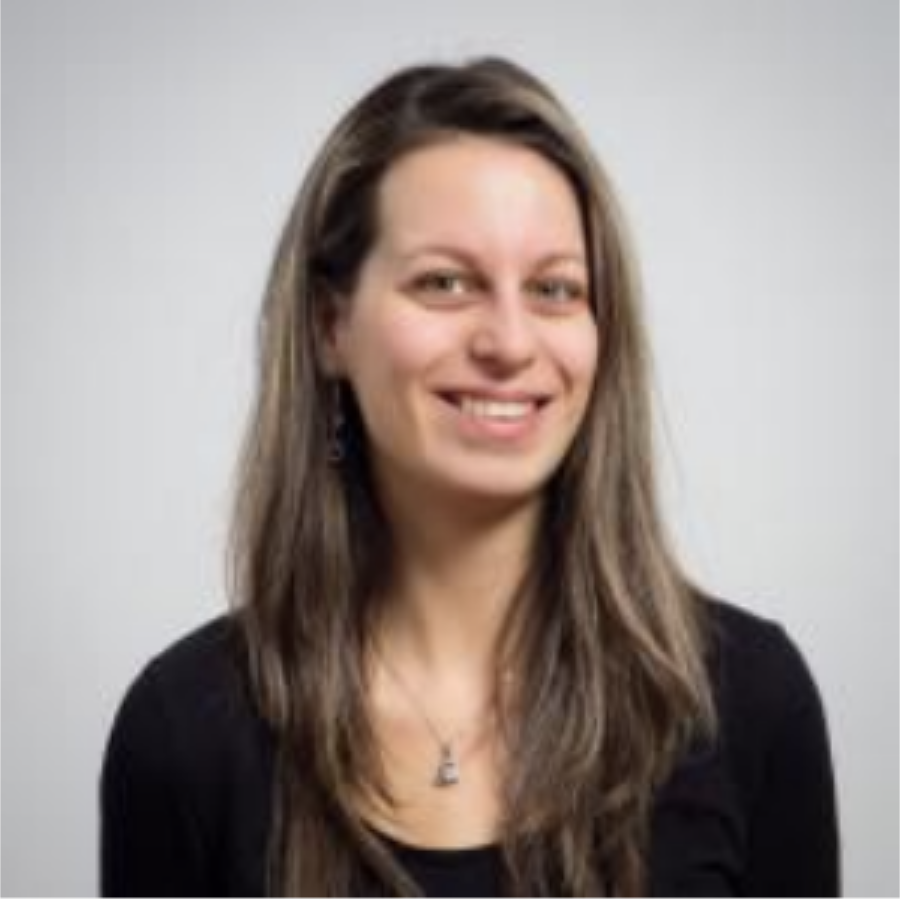}}Michaela,Bačíková\\ is an assistant professor and the Head of the Information Systems Laboratory at the Department of Computers and Informatics, Technical University of Košice, Slovakia. She received her Ph.D. in Computer Science in 2014. Since 2014 she is a member of the Department of Computers and Informatics at Technical University of Košice. Currently, the main subject of her research is UX, HCI, and usability while focusing on the domain-related terminology in user interfaces (domain usability). She also focuses on software languages and innovations in the teaching process.

\bio{\includegraphics[width=0.22\textwidth]{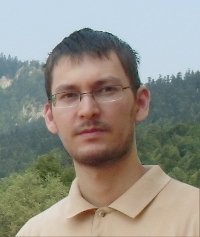}}Matúš,Sulír\\ is an assistant professor at the Department of Computers and Informatics, Technical University of Košice, Slovakia. At the same university, he graduated with a Master's degree in Computer Science in 2014 and Ph.D. in 2018. His research is focused on program comprehension, particularly on the integration of run-time information with source code, attribute-oriented programming, and debugging. He is also interested in empirical studies in software engineering.

\bio{\includegraphics[width=0.22\textwidth]{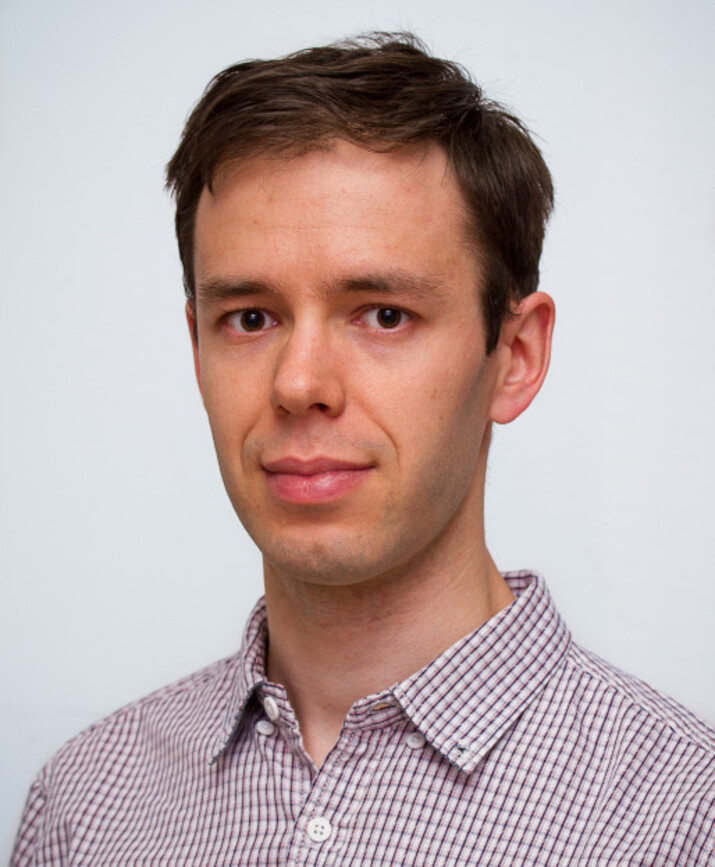}}Jan,Juhár\\ is a researcher at the Department of Computers and Informatics, Technical University of Košice. He received his Ph.D. in Computer Science in 2018. Since 2018 he is a member of the Department of Computers and Informatics at Technical University of Košice. His research focuses on program comprehension, programming tools, source code metadata, and program projections.

\bio{\includegraphics[width=0.22\textwidth]{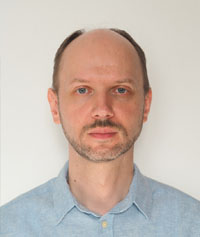}}Sergej,Chodarev\\ is an assistant professor at the Department of Computers and Informatics, Technical University of Košice, Slovakia. He received his Master's degree in computer science in 2009 and his Ph.D. degree in computer science in 2012. The subject of his research includes domain-specific languages, metaprogramming, and software engineering.

\bio{\includegraphics[width=0.22\textwidth]{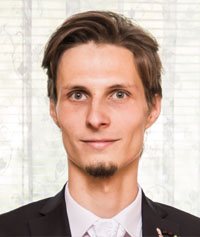}}Filip,Gurbáľ\\ is a Ph.D. student at the Department of Computers and Informatics, Technical University of Košice, Slovakia. He received his Ing. in Computer Science in 2020. He is a member of the Computer Network Laboratory at the Technical University of Košice. The subject of his research is improving program comprehension using methods and tools. He also focuses on software testing methods and tools.

\label{lastpage}
\end{document}